\documentclass{article}
\usepackage{spconf,amsmath,graphicx}
\usepackage{amsfonts} 
\usepackage{IEEEtrantools} 
 
\usepackage{epstopdf}  
\usepackage{verbatim}  
\usepackage{color} 
\usepackage{pifont} 
\usepackage{pmat} 
\usepackage{booktabs} 
\usepackage{esvect}  
\usepackage{pgfplots}  
\usetikzlibrary{plotmarks}
\usepackage{balance}

\DeclareMathOperator{\Diag}{\mathsf{bdiag}}
\DeclareMathOperator{\Stack}{\mathsf{stack}}

\DeclareMathOperator{\Vn}{\mathbf{V}}

\DeclareMathOperator{\hn}{\underline{\mathbf{h}}}
\DeclareMathOperator{\hen}{\hat{\mathbf{h}}}
\DeclareMathOperator{\hern}{\tilde{\mathbf{h}}}
\DeclareMathOperator{\Hn}{\mathbf{H}}
\DeclareMathOperator{\Hneq}{\mathbf{\bar H}}
\DeclareMathOperator{\0}{\mathbf{0}}
\DeclareMathOperator{\I}{\mathbf{I}}

\newcommand{\Cmat}[2]{\!\in\mathbb{C}^{#1\times#2}}

\newcommand{\textbox}[3]{\node[draw=blue,thick,fill=white, font=\scriptsize, minimum size = 5mm, rounded corners] at (axis cs: #1,#2) {#3};}  
         
\newcommand{\textboxy}[3]{\node[draw=red,dashed,thick,fill=white, font=\scriptsize, minimum size = 5mm, rounded corners] at (axis cs: #1,#2) {#3};}

\renewcommand{\IEEEQED}{\IEEEQEDopen} 
\renewcommand{\IEEEproofindentspace}{0em}

\newtheorem{theorem}{Theorem}
\newtheorem{corollary}{Corollary}

\title{On the Degrees of freedom of the $\textit{\textbf{K}}$-user MISO \\
Interference Channel with imperfect delayed CSIT}
\name{Marc Torrellas, Adrian Agustin, Josep Vidal\thanks{This work has been done in the framework of the projects TEC2010-19171/TCM and CONSOLIDER INGENIO CSD2008-00010 COMONSENS, and by project 2009SGR1236
(AGAUR) of the Catalan Administration, and TROPIC FP7 ICT-2011-8-318784 project, funded by the European Commission. Draft version of the accepted manuscript at IEEE ICASSP 14.}}
\address{Universitat Polit\`ecnica de Catalunya (UPC), Barcelona \\
\{marc.torrellas.socastro, adrian.agustin, josep.vidal\}@upc.edu}
\begin{document}

\copyright 2014 IEEE. Personal use of this material is permitted. Permission from IEEE must be 
obtained for all other uses, in any current or future media, including 
reprinting/republishing this material for advertising or promotional purposes, creating new 
collective works, for resale or redistribution to servers or lists, or reuse of any copyrighted 
component of this work in other works.

\balance
\newpage

\maketitle
\begin{abstract}
This work investigates the degrees of freedom (DoF) of the $K$-user multiple-input single-output (MISO) interference channel (IC) with imperfect delayed channel state information at the transmitters (dCSIT). For this setting, new DoF inner bounds are provided, and benchmarked with cooperation-based outer bounds. The achievability result is based on a precoding scheme that aligns the interfering received signals through time, exploiting the concept of \textit{Retrospective Interference Alignment} (RIA). The proposed approach outperforms all previous known schemes. Furthermore, we study the proposed scheme under channel estimation errors (CEE) on the reported dCSIT, and derive a closed-form expression for the achievable DoF with imperfect dCSIT.
\end{abstract}
\section{Introduction}
\label{sec:intro}

 
The characterization of wireless networks in DoF terms has attracted researchers world-wide during the last years \cite{Maddah-Ali2008,CJ}. In this context, the \textit{interference alignment} (IA) concept is one of the main design tools for DoF-optimal communication strategies in interference-based networks \cite{CJ,SubspaceAlignmentChains,GenieChains}. 
Originally, IA-based techniques were developed assuming perfect and current CSIT. Nevertheless, this assumption is in general too optimistic. Recently, \cite{MAT} has investigated the $K$-user MISO Broadcast Channel with $M\geq K$ antennas at the transmitter and perfect delayed CSIT (dCSIT), i.e. perfect CSI of previous time slots, but no current CSI knowledge. There is shown that even with completely outdated CSIT, the number of DoF is larger than in the no CSIT case \cite{VazeNoCSI_Tr}. The scheme consists on a multi-phase transmission where signals are designed in such a way that the received interfering signals are aligned along the space-time domain. For this reason, the concept behind this strategy was referenced as RIA in \cite{MalekiRIA}.
 
A lot of interest has come up for analyzing the DoF of the IC with dCSIT, see for example \cite{MalekiRIA,Vaze2IC,Ghasemi2011,Abdoli_IC,3UMIC}, using a similar approach as the one proposed in \cite{MAT}. All these works extend the same principle to the IC, where each transmitter only transmits to its intended receiver. It is specifically related to this work the best known result for the $K$-user MISO IC with local dCSIT, presented in \cite{Ghasemi2011}, or the work in \cite{3UMIC} where the DoF are studied for $K=3$ users and multi-antenna receivers.

The impact of imperfect CSI on the DoF has also been considered in the literature. For the case of imperfect current CSIT, it is known that the full multiplexing gain can be obtained as long as the feedback is reported with an uplink SNR comparable to the downlink SNR \cite{chEstErrorPCSIT,IAanalogCSI}. This analysis has been extended for the case of imperfect dCSIT in \cite{ChElia_Journal}, where the authors analyze the Broadcast Channel and show that the optimal DoF region under perfect dCSIT can be obtained as of a minimum channel feedback quality (FBQ) value.
 
This work analyzes the $K$-user MISO IC, where transmitters are equipped with $M \geq K$ antennas and have imperfect local dCSIT, i.e. each transmitter estimates the CSI of the channels departing from this transmitter during previous phases. The main contributions of this work are: 
\begin{enumerate}
\item We develop new inner and outer bounds for the DoF of the $K$-user MISO IC with imperfect local dCSIT, improving previously known results when particularizing to the perfect dCSIT case. Theorem 1 in section \ref{sec:MainResults} presents a closed-form expression for our DoF results. 
\item We propose a simple precoding scheme, applicable to any number of users. Our approach addresses the $K$-user MISO IC with $M \!\!\! \geq \!\!\! K$ antennas at the transmitters and single-antenna receivers, but it can easily be adapted to a MIMO scenario where users have $N$ antennas and transmitters have $M \geq KN$.
\end{enumerate}
$\mathbf{Notation}$: Boldface and lower case fonts denote column vectors ($\mathbf{x}$), while row vectors are also underlined ($\underline{\mathbf{x}}$). Boldface and upper case is used for matrices ($\mathbf{X}$). $(·)^H$, $\0$, and $\I$ are the transpose and conjugate operator, the all-zero matrix, and the identity matrix, respectively. Also, we use the next matrix operations:
\begin{IEEEeqnarray*}{c}
 \begin{matrix}
\Stack
\left( \mathbf{A},\mathbf{B}\right) =
\begin{pmatrix}
\mathbf{A} \\
\mathbf{B} \\
\end{pmatrix} 
&
\Diag
\left( \mathbf{A},\mathbf{B} \right) =
\begin{pmatrix}
	\mathbf{A} & \0 \\
	\0 & \mathbf{B} \\
\end{pmatrix}
\end{matrix}
\end{IEEEeqnarray*}

\begin{figure}[t]
\begin{minipage}[b]{1\linewidth}
  \centering    
  \centerline{\includegraphics[width=0.6\linewidth]{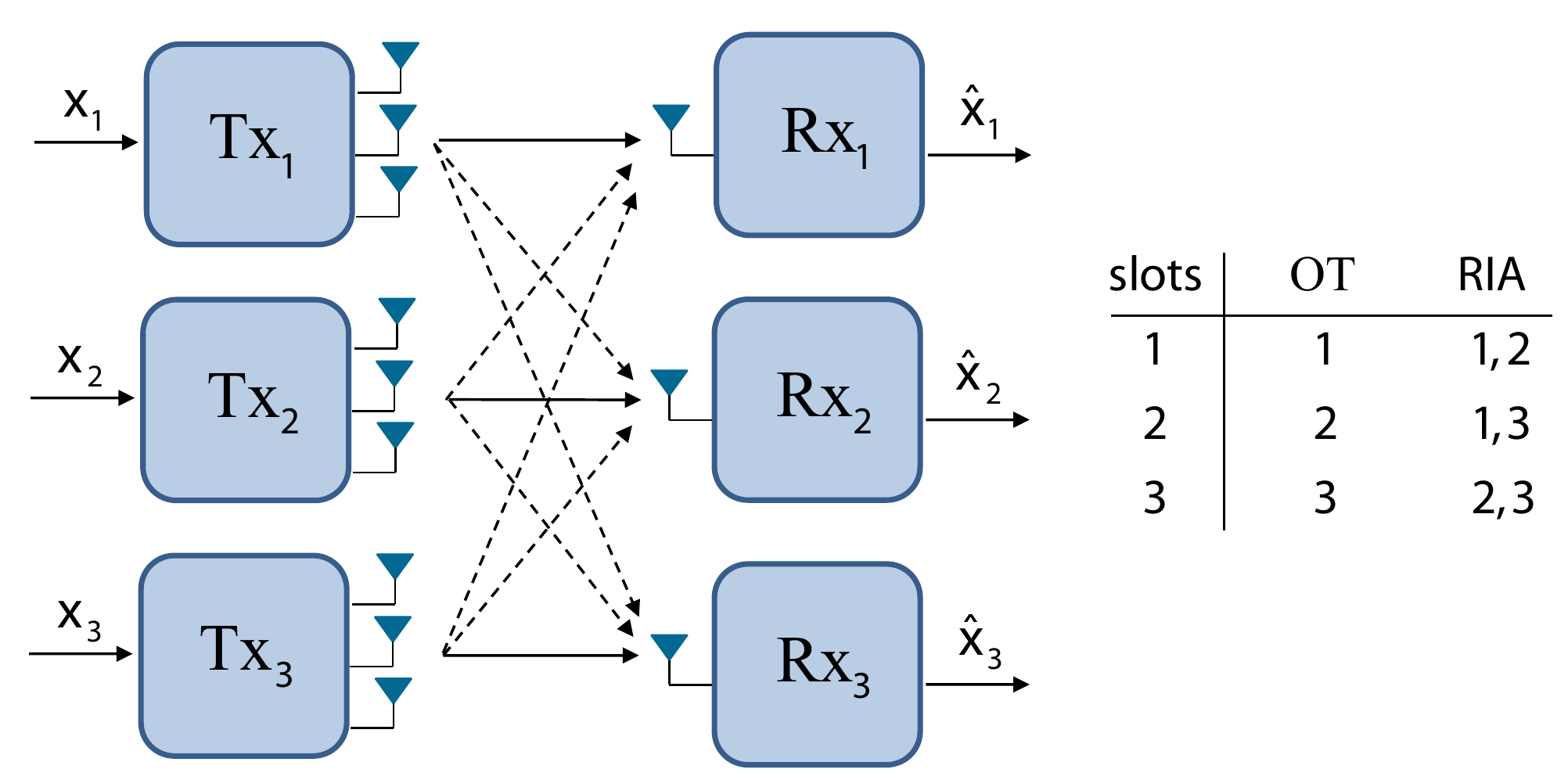}}
  
\end{minipage} 
%
\caption{  The $3$-user MISO IC with $(3,1)$ antennas. The transmission is carried out in 6 slots, divided in two phases: OT and RIA. The right hand side table shows the active transmitters during each time slot.}
\label{fig:scenarios}
\end{figure}
\section{System Model}
\label{sec:SystemModel}
We consider a flat fading interfering channel with $K$ transmitter/receiver pairs, where transmitters and receivers are equipped with $({M \geq K},1)$ antennas, respectively. The transmission is carried out in 
$W$ time slots, divided in $2$ phases of duration $W_1$ and $W_2$ slots, respectively, where each transmitter aims to deliver $b$ symbols to its associated receiver. During the first phase (\textit{orthogonal transmission (OT) phase}) only one transmitter is active per slot, whereas for the second phase (\textit{RIA phase}), users act by pairs. The total number of slots for each phase is given by $W_p=\binom{K}{p}$ slots, i.e. the number of possible groups of $p$ users out of the total $K$ users. Fig. \ref{fig:scenarios} depicts the transmission protocol for the $K=3$ case. The aim of the OT phase is to help users sensing the interfering transmitters. That gathered information is then used during the RIA phase to align the generated interference to non-intended receivers with the signals received during the first phase. Further details will be given in Section \ref{sec:KUMIC}.  

We refer to the time instant $(p,s)$ as the $s$th time slot of the $p$th phase, and the set $G_{(p,s)}$ specifies the active transmitters during the $(p,s)$th time slot. The output at the $j$th receiver for the time instant $(p,s)$ is described by:
\begin{IEEEeqnarray}{c}
y_j^{\left(p,s\right)} = 
\sum_{i \in G_{(p,s)}} \hn_{j,i}^{\left(p,s\right)} \Vn_i^{\left(p,s\right)}\mathbf{x}_i + n_j^{(p,s)}
\label{eq:SystemModel}
\end{IEEEeqnarray}
where $\Vn_i^{\left(p,s\right)}\! \Cmat{M}{b}$ is the precoding matrix used by the $i$th user at time slot $(p,s)$, $\mathbf{x}_i \! \Cmat{b}{1}$ contains the zero-mean unit-variance i.i.d. symbols intended to be decoded at the $i$th receiver, $n_j^{(p,s)}$ is the zero-mean unit-variance AWGN term, and $\hn_{j,i}^{\left(p,s\right)}\! \Cmat{1}{M}$ is the channel vector during the time slot $(p,s)$ containing the channel gains from antennas of the $i$th transmitter to the $j$th receiver.
Note that ${\Vn_i^{(p,s)}=\0}, \forall i \notin G_{(p,s)}$, and we force $
{\mathbb{E} \big\{\big\| 
    \Vn_i^{\left(p,s\right)} \! \mathbf{x}_i
\big\|^2\big\} \leq P
}$, where $P$ represents in our case the signal to noise ratio (SNR).   

The reported CSI is estimated with finite accuracy, assuming the CSI is fed back with an uplink SNR equal to $ P^{\epsilon}$, where $\epsilon \! \in \! [0,1]$ captures the FBQ. Hence, $\hn_{j,i}^{\left(p,s\right)}$ is estimated by $\hat \hn_{j,i}^{\left(p,s\right)}$, and the channel estimation error (CEE) is given by
\begin{IEEEeqnarray}{c}
\tilde\hn_{j,i}^{\left(p,s\right)} = \hn_{j,i}^{\left(p,s\right)} \! - \, \hat{\hn}_{j,i}^{\left(p,s\right)},\text{ with }  \,
\mathbb{E} \big\{\left\| 
    \tilde\hn_{j,i}^{\left(p,s\right)} 
\right\|^2\big\} = P^{-\epsilon}
\label{eq:estimationError}
\end{IEEEeqnarray}
where $\epsilon=0$ stands for totally distorted dCSIT, since the CEE power does not vanish when $P \to \infty$, and distorts the channel observation at the transmitters. On the other hand, when $\epsilon\!=\!1$ the dCSIT is as good as perfect, since the power of the CEE goes to zero when $P$ is sufficiently high, and thus has no impact in the DoF sense. Note that the FBQ characterization described above is valid only for the high SNR regime, because otherwise, $ \epsilon =\!1$ cannot be assumed as perfect dCSIT.

We assume local imperfect dCSIT knowledge. According to previous definitions, this means that transmitter $j$ has access to $\{ \hen_{1,j}^{\left(1,s\right)}, \hen_{2,j}^{\left(1,s\right)} \ldots 
\hen_{K,j}^{\left(1,s\right)} \}$ at the beginning of the RIA phase, for every slot $s$ where $j \in G_{(1,s)} $, i.e. each time slot of the first phase where the $j$th transmitter is active.


After the two phases, all the received signals can be grouped as follows:
\begin{IEEEeqnarray}{c}
\label{eq:SystemModelextended}
\begin{matrix}
\mathbf{r}_j = \mathbf{U}_j \left(
\sum_{i=1}^{K} \mathbf{H}_{j,i} \mathbf{V}_i \mathbf{x}_i + \mathbf{n}_j \right) \\
\mathbf{H}_{j,i} = \Diag
\left(
\begin{matrix}
	{\hn_{j,i}^{\left(1,1\right)}} &
 	{\hn_{j,i}^{\left(1,2\right)}} & \ldots &
	{\hn_{j,i}^{\left(2,W_2 \right)}}
\end{matrix} 
\right)
\\
 \mathbf{V}_i = \Stack
\left(
\begin{matrix}
	{\Vn_i^{\left(1,1\right)}}
	&{\Vn_i^{\left(1,2\right)}} 
	& \ldots & {\Vn_i^{\left(2,W_2\right)}}
\end{matrix}
\right)  
\end{matrix}
\end{IEEEeqnarray}
where ${\mathbf{V}_i\Cmat{MW}{b}}$ is the global $i$th precoding matrix, ${\mathbf{H}_{j,i} \Cmat{W}{MW}}$ is the extended $j$th channel matrix, ${\mathbf{r}_j \Cmat{b}{b}}$ is the signal processed at the $j$th receiver, and ${\mathbf{U}_j\Cmat{b}{W}}$ is a $j$th linear filter devoted to combat the interference.
For the proper DoF analysis in section \ref{sec:DoFanalysis}, let define the matrix ${\mathbf{\Xi}_{j} \Cmat{W}{(K-1)b}}$ as a matrix containing the received interference signals during the whole transmission. This matrix is computed by concatenating the columns $\Hn_{j,i} \Vn_i$, ${\forall i \neq j}$. Finally, the achieved DoF per user are defined as
\begin{IEEEeqnarray}{c}
\hat{d_j} \! = \lim_{P \to \infty} \! \! \frac{\mathbb{E} \left \{ I \left( \mathbf{x}_j ; \mathbf{r}_j \right) \right\}}{W \log P}
\label{eq:DoFdef}
\end{IEEEeqnarray} 
where $I(a;b)$ stands for the mutual information of $a$ and $b$.
          
\section{Main result}
\label{sec:MainResults} 
  
The main contribution of this work is next stated:

\begin{theorem}
The DoF per user of the $K$-user MISO IC with imperfect local dCSIT, with FBQ $0 \leq \epsilon \leq 1$, and with ${(M \ge K,1)}$ antennas at the transmitters and receivers, respectively, is given by
\begin{IEEEeqnarray}{c}
\frac{2}{K+1}\frac{1+(K-1)\epsilon}{K} \leq d \leq \left({\sum_{i=1}^K \frac{1}{i}}\right)^{-1}
\label{eq:DoFthereom}
\end{IEEEeqnarray} 
 \label{th:KMISOIC} 
\end{theorem}
\begin{IEEEproof}
See section \ref{sec:KUMIC} for the inner bound proof. The outer bound relies on transmitter cooperation, i.e. a BC with dCSIT where the transmitter is equipped with $KM$ antennas \cite{MAT}.
\end{IEEEproof}

\begin{corollary}
 For a $K$-user MIMO IC with $(M \geq KN,N)$, the proposed DoF inner and outer bounds are scaled by $N$.
\end{corollary}

\begin{figure}[t] 
\begin{minipage}[t]{1\linewidth}
  \centering
%
%
%
%

\definecolor{mycolor1}{rgb}{1,0,1}

\begin{tikzpicture}

\begin{axis}[%
view={0}{90},
width=0.5\linewidth,
scale only axis,
xmin=2, xmax=10,
xlabel={users},
xmajorgrids,
ymin=0, ymax=1,
ylabel={DoF per user},
ylabel style={at={(0.04,0.5)}},
ymajorgrids,
grid style={dashed},
legend style={at={(1,0.81)},align=left}]
\addplot [
color=blue,
solid,
thick,
mark size=3pt,
mark=triangle,
mark options={solid,,rotate=180}
]
coordinates{
 (1,0)(2,0.666666666666667)(3,0.545454545454546)(4,0.48)(5,0.437956204379562)(6,0.408163265306123)(7,0.385674931129477)(8,0.367936925098555)(9,0.353485762379015)(10,0.341417152147406) 
};
\addlegendentry{outer bound Theorem 1};

\addplot [
color=red,
dashed,
thick,
mark size=3pt,
mark=triangle,
mark options={solid,,rotate=180}
]
coordinates{
 (1,1)(2,0.666666666666667)(3,0.714285714285714)(4,0.769230769230769)(5,0.80952380952381)(6,0.838709677419355)(7,0.86046511627907)(8,0.87719298245614)(9,0.89041095890411)(10,0.901098901098901) 
};
\addlegendentry{outer bound \cite{Ghasemi2011}};

\addplot [
color=blue,
solid,
thick,
mark size=3pt,
mark=triangle,
mark options={solid}
]
coordinates{
 (1,1)(2,0.666666666666667)(3,0.5)(4,0.4)(5,0.333333333333333)(6,0.285714285714286)(7,0.25)(8,0.222222222222222)(9,0.2)(10,0.181818181818182) 
};
\addlegendentry{inner bound Theorem 1};

\addplot [
color=red,
dashed,
thick,
mark size=3pt,
mark=triangle,
mark options={solid,draw=red}
]
coordinates{
 (1,1)(2,0.666666666666667)(3,0.428571428571429)(4,0.307692307692308)(5,0.238095238095238)(6,0.193548387096774)(7,0.162790697674419)(8,0.140350877192982)(9,0.123287671232877)(10,0.10989010989011) 
};
\addlegendentry{inner bound \cite{Ghasemi2011}};

\addplot [
color=black,
solid,
thick,
mark size=2pt,
mark=square,
mark options={solid}
]
coordinates{
 (2,0.5)(3,0.387096774193548)(4,0.296052631578947)(5,0.239111870196413)(6,0.200596056854654)(7,0.172832321888769)(8,0.151863704422931)(9,0.135460984743207)(10,0.122274614750177) 
};
\addlegendentry{inner bound SISO IC \cite{Abdoli_IC}};

\addplot [
color=mycolor1,
solid,
thick,
mark size=1.5pt,
mark=o,
mark options={solid}
]
coordinates{
 (2,0.5)(3,0.333333333333333)(4,0.25)(5,0.2)(6,0.166666666666667)(7,0.142857142857143)(8,0.125)(9,0.111111111111111)(10,0.1) 
};
\addlegendentry{TDMA (no CSIT)};

\end{axis}
\end{tikzpicture}%
\end{minipage}
\caption{DoF per user inner and outer bounds for the $K$-user MISO IC with ${M \geq K}$ antennas at each transmitter and with perfect dCSIT ($\epsilon=1$).}
\label{fig:IC_ghasemi_us}
\end{figure}
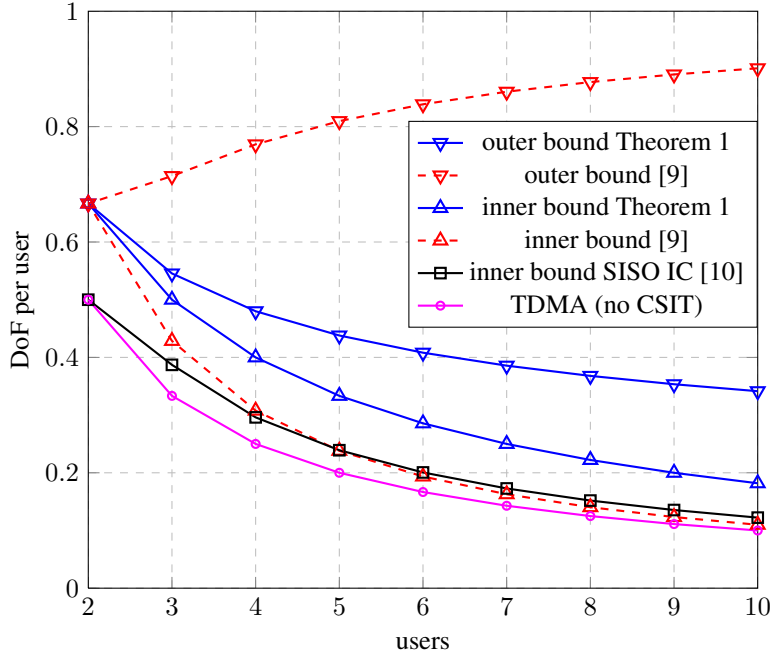

Fig. \ref{fig:IC_ghasemi_us} depicts the obtained DoF for perfect dCSIT (${\epsilon=1}$). Results are compared with the scheme introduced in \cite{Ghasemi2011}, with TDMA\footnote{TDMA achieves the optimal DoF when no CSIT is assumed \cite{VazeNoCSI_Tr}.}), and with the scheme proposed in \cite{Abdoli_IC} for the $K$-user SISO IC assuming $\text{global}$ dCSIT. 
Some interesting remarks are next commented:

\begin{itemize}
 \item Our inner and outer bounds are tighter than any previously reported results.
 \item In case of perfect dCSIT ($\epsilon \! = \!1$), and as the number of users increases, the scheme in \cite{Ghasemi2011} collapses to the TDMA performance, whereas our scheme achieves twice the corresponding DoF.
 \item The proposed scheme outperforms TDMA whenever $\epsilon >  \frac{1}{2}$, see (\ref{eq:DoFthereom}), regardless the number of users $K$. 
\end{itemize}
    
%
%
%
%

\section{Transmission Protocol}
\label{sec:KUMIC}

%
%
%

This section derives a transmission scheme able to achieve $ \frac{2}{K+1}\frac{1+(K-1)\epsilon}{K} $ DoF per user. Our approach delivers $b = K $ symbols to each of the $K$ users along $W = K + \binom{K}{2}$ slots. For simplicity, we consider the $M=K$ case\begin{footnote}{It may be expected a {\it diversity gain} when $M \geq K$, but the interest of this paper is focused on the {\it multiplexing gain} (DoF).}\end{footnote}, while the case $M \geq K$ can be tackled by turning off the additional antennas. The scheme is divided in two phases: OT and RIA phases.

\subsection{Orthogonal Transmission phase}
\label{IC-p1}

During this phase, users are served in TDMA, i.e. ${G_{(1,s)}=\{ s \} }$. The signal received at each receiver during each time slot is given by:
\begin{IEEEeqnarray}{c}
y_j^{\left(1,s\right)} = 
\hn_{j,s}^{\left(1,s\right)} \Vn_s^{\left(1,s\right)}\mathbf{x}_s + n_j^{(1,s)}
\label{eq:SystemModel_fase1}
\end{IEEEeqnarray}
Since there is no CSIT, one symbol is transmitted with no precoding per antenna and slot, i.e. $\Vn_s^{(1,s)} = \sqrt{\frac{P}{K}} \I, {s=1\ldots W_1}$. Consequently, each user gets $1$ linear combination (LC) of its $K$ desired symbols, thus it is not able to decode them yet. 
On the other hand, $1$ LC of each interfering signal is overheard at each receiver. This overheard information will be exploited by the transmitters in the RIA phase.

\subsection{Retrospective Interference Alignment phase}
\label{sec:IC-p2}

During the second phase, each transmitter is active in ${K-1}$ slots, transmitting simultaneously with another transmitter, obtained from the $\binom{K}{2}$ possible pairs. Let us assume that in the $s$th slot the two transmitters $\alpha$ and $\beta$ are active, i.e. ${G_{(2,s)}=\left\{\alpha, \beta \right\}}$, and the received signal at receiver $\beta$ is,
\begin{IEEEeqnarray}{c}
y_{\beta}^{\left(2,s\right)} = 
{\hn_{\beta,\beta}^{\left(2,s\right)}} \!\!			 
    \Vn_{\beta}^{\left(2,s\right)}\mathbf{x}_{\beta} + 
{\hn_{\beta,\alpha}^{\left(2,s\right)}} \!\! 
    \Vn_{\alpha}^{\left(2,s\right)}\mathbf{x}_{\alpha} + n_{\beta}^{(2,s)}
\label{eq:SystemModel_fase2}
\end{IEEEeqnarray}
Then, the precoding matrix $\mathbf{V}_{\alpha}^{(2,s)}$ is designed such that the interference created at the $\beta$th  receiver is aligned with the previous received interference. 
The RIA condition at this receiver for this time slot is expressed by
\begin{IEEEeqnarray}{c}
 {\hn_{\beta,\alpha}^{(2,s)}} \! \mathbf{V}_{\alpha}^{(2,s)} \propto {\hn_{\beta,\alpha}^{(1,\alpha)}} \! \mathbf{V}_{\alpha}^{(1,\alpha)}
\label{eq:IC-M3p2_design}
\end{IEEEeqnarray}
and, since there is no current CSIT, we select
\begin{IEEEeqnarray}{c}
\mathbf{V}_{\alpha}^{(2,s)} = 
{\sigma}_{\alpha}^s \begin{bmatrix} 1 \\ \0 \end{bmatrix} \hat\hn_{\beta,\alpha}^{(1,\alpha)}
\label{eq:IC-M3p2_Vsolution}
\end{IEEEeqnarray} 
where ${\sigma}_{\alpha}^s$ ensures the transmit power constraint. 

\subsection{DoF performance}
\label{sec:DoFanalysis}

Now we analyze the DoF achieved for each user subject to the imperfect dCSIT assumption. Due to space limitation, only a sketch of the proof for the $K=3$ case is shown, whereas the generalization to $K$ users is straightforward. In this regard, consider the matrix of interference at user 1:
\begin{IEEEeqnarray}{c}
\mathbf{\Xi}_1 =
\begin{bmatrix}
 \0 & \0 \\
 c_1 \sqrt{P}  {\hn_{1,2}^{(1,2)}} & \0 \\
 \0 & c_1 \sqrt{P}  {\hn_{1,3}^{(1,3)}} \\
 c_2 \sqrt{P} {\hen_{1,2}^{(1,2)}} & \0 \\
 \0 & c_3 \sqrt{P}  {\hen_{1,3}^{(1,3)}}  \\
 c_4 \sqrt{P}  {\hen_{3,2}^{(1,2)}} & c_5 \sqrt{P} {\hen_{2,3}^{(1,3)}}
\end{bmatrix}
\label{eq:interference}
\end{IEEEeqnarray}
where constants $c_k, k=1 \ldots 5$ are independent of $P$, and designed to satisfy the transmitted power constraint. Remember that each row in (\ref{eq:interference}) corresponds to each time slot, and each block column to users 2 and 3, respectively.

The receiver aims to suppressing the interference by using the received signal in different phases, hence user 1 applies the following receive filter,
\begin{IEEEeqnarray}{c} 
\mathbf{U}_1 =
\begin{bmatrix}
 1 & 0 & 0 & 0 & 0 & 0  \\
 0 & c_2 & 0 & -c_1 & 0 & 0  \\
 0 & 0 & c_3 & 0 & -c_1 & 0  
\end{bmatrix}
\label{eq:receiverImperfect}
\end{IEEEeqnarray}
and the residual interference at receiver 1 results,
\begin{IEEEeqnarray}{c} 
\mathbf{U}_1 \mathbf{\Xi}_1 =
\begin{bmatrix}
 \0 & \0  \\
 c_1 c_2 \sqrt{P} \hern_{1,2}^{(1,2)}  & \0  \\
 \0 & c_1 c_3 \sqrt{P} \hern_{1,3}^{(1,3)}
\end{bmatrix}
\label{eq:residualInterf}
\end{IEEEeqnarray}
with $\hern_{j,i}^{(p,s)}$ defined as in (\ref{eq:estimationError}). Therefore, the interference plus noise covariance matrix is given by,
\begin{IEEEeqnarray}{c} 
{\mathbf{\Upsilon}_j = \mathbf{U}_j {\mathbf{\Xi}}_j {\mathbf{\Xi}}_j^H \mathbf{U}_j^H + \mathbf{U}_j \mathbf{U}_j^H} \\
\mathbf{\Upsilon}_j = 
\begin{bmatrix}
 1 & 0  & 0 \\
 0 & c_6 P \, \big \| \! \hern_{1,2}^{(1,2)} \! \big \|^2  & 0  \\
 0 & 0 & c_7 P \, \big \| \! \hern_{1,3}^{(1,3)} \! \big \|^2
\end{bmatrix}
\label{eq:covInterf}
\end{IEEEeqnarray}
where $c_6, c_7$ are independent of $P$, and we assume $P \to \infty$. Finally, we obtain the achieved DoF by using (\ref{eq:DoFdef}), as follows,
\begin{align}
  \hat{d}_1 &{\geq}  
  \frac{\log \mathbb{E} \left \{ \left| \Hneq_1 \Hneq_1^H \right| \right \} - \log \mathbb{E} \left\{ \left| \mathbf{\Upsilon}_1 \right| \right \}}{6\log P}  \label{eq:DoFimperfect1}\\
& = 
  \frac{3\log{P} - 2 (1-\epsilon)\log P}{6 \log P} 
= 
\frac{1+2\epsilon}{6}
\label{eq:DoFimperfect2}
\end{align}
where $\Hneq_{j}=\mathbf{U}_{j} \mathbf{{H}}_{j,j} \mathbf{{V}}_j$ is the $j$th equivalent channel, and (\ref{eq:DoFimperfect1})-(\ref{eq:DoFimperfect2}) follow assuming $P \to \infty$, and using the Jensen's inequality \cite{Boyd2004} and basic properties of linear algebra. Due to symmetry of the problem, the same arguments hold for users 2 and 3. The DoF value stated in Theorem 1 is obtained by generalizing this procedure to the $K$-user case.

\section{Simulation Results}
For the 3-user case, we evaluate our scheme for different FBQ and SNR values. Fig.\ref{fig:sims} shows the rate per user as a function of the SNR, averaged over 2000 channel realizations. We compare our results with the no CSIT case, referring to developing a TDMA strategy during 9 time slots. Notice that the slope for $\epsilon=0.5$ coincides with that for the no CSIT case, as expected from the DoF expression derived in Theorem 1. This implies that the dCSIT scheme outperforms the no CSIT case as long as the uplink SNR is higher than half the downlink SNR (both in dB). It is also remarkable that the no CSIT scheme can be outperformed with the sufficiently high $\epsilon$, even in the low-medium SNR regime. Further, Fig.~\ref{fig:sims2} presents the 10-percentile outage rate per user as a function of FBQ ($\epsilon$). It can be seen that the required FBQ value for outperforming the TDMA scheme depends on the SNR.

\section{Conclusions}
\label{sec:conclusions}

The DoF of the $K$-user MISO IC have been studied when the transmitters have imperfect, delayed, and local CSIT. We propose a simple precoding scheme and analyze its performance in terms of DoF as a function of the Feedback Quality $\epsilon$. This DoF inner bound is contextualized with a new DoF outer bound, both improving all previous results. Additionally, we find out that as long as $\epsilon>\frac{1}{2}$ the RIA scheme outperforms the no CSIT case. Finally, simulation results validate the theoretical analysis, and show the benefits of using dCSIT not only in terms of average rate but also in terms of outage rate with respect to the case of uninformed transmitters.
    
\begin{figure}[h] 
\begin{minipage}[]{1\linewidth}   
  \centering    
%
%
%
%

\definecolor{mycolor1}{rgb}{1,0,1}

\begin{tikzpicture}

\begin{axis}[%
view={-0}{90},
width=0.5\linewidth,
scale only axis,
xmin=5, xmax=40,
xlabel={SNR (dB)},
xmajorgrids,
ymin=0.5, ymax=6,
ylabel={Rate per user (bps/Hz)},
ylabel style={at={(0.06,0.5)}},
ymajorgrids,
axis lines*=left,
grid style={dashed},
legend style={at={(0.03,0.97)},anchor=north west,align=left}]
\addplot [
color=red,
solid,
thick,
mark=square,
mark options={solid}
]
coordinates{
 (5,0.634524473780716)(10,1.0974009120309)(15,1.6812462453344)(20,2.35798085127099)(25,3.10031499937658)(30,3.88444945291215)(35,4.69240072077191)(40,5.5125940593238) 
};
\addlegendentry{$\epsilon=1$};

\addplot [
color=mycolor1,
dash pattern=on 1pt off 3pt on 3pt off 3pt,
thick,
mark=square,
mark options={solid}
]
coordinates{
 (5,0.611694405833348)(10,1.06429257587745)(15,1.636214876025)(20,2.30000273964627)(25,3.029230065703)(30,3.80073184527609)(35,4.59672944036626)(40,5.40541791523706) 
};
\addlegendentry{$\epsilon\text{ = 0.9}$};

\addplot [
color=blue,
dashed,
thick,
mark=square,
mark options={solid}
]
coordinates{
 (5,0.605561915239053)(10,1.04628330440784)(15,1.59680216178491)(20,2.22750056005023)(25,2.91151592342009)(30,3.62657380713162)(35,4.35602692304969)(40,5.08901475128656) 
};
\addlegendentry{$\epsilon\text{ = 0.7}$};

\addplot [
color=red,
solid,
thick,
mark=o,
mark options={solid}
]
coordinates{
 (5,0.598342927472829)(10,1.02164726215216)(15,1.53556643658253)(20,2.10387702392623)(25,2.69735409888195)(30,3.29680946293836)(35,3.89195231252071)(40,4.4789281169969) 
};
\addlegendentry{$\epsilon\text{ = 0.5}$};


\addplot [
color=mycolor1,
dash pattern=on 1pt off 3pt on 3pt off 3pt,
thick,
mark=o,
mark options={solid}
]
coordinates{
 (5,0.585826913215771)(10,0.974398356299346)(15,1.41169919882943)(20,1.85117760561897)(25,2.26882363325092)(30,2.66172816408375)(35,3.03649032766681)(40,3.4014507265861) 
};
\addlegendentry{$\epsilon\text{ = 0.2}$};


\addplot [
color=blue,
dashed,
thick,
mark=triangle,
mark options={solid,,rotate=180}
]
coordinates{
 (5,0.577099630812466)(10,0.943237018666566)(15,1.33473263051758)(20,1.7060587090084)(25,2.04289379401255)(30,2.35194062009064)(35,2.64460762518311)(40,2.92921653335922) 
};
\addlegendentry{$\epsilon\text{ = 0.01}$};

\addplot [
color=green,
solid,
thick
]
coordinates{
 (5,0.571038862252191)(10,0.967053596419586)(15,1.44096074005562)(20,1.95859023738253)(25,2.49724868779673)(30,3.04501955766741)(35,3.59646224836546)(40,4.14930578057426) 
};
\addlegendentry{no CSIT};


\end{axis}
\end{tikzpicture}%

\end{minipage}
\caption{Average rate per user vs SNR for different FBQ}
\label{fig:sims}
\end{figure}
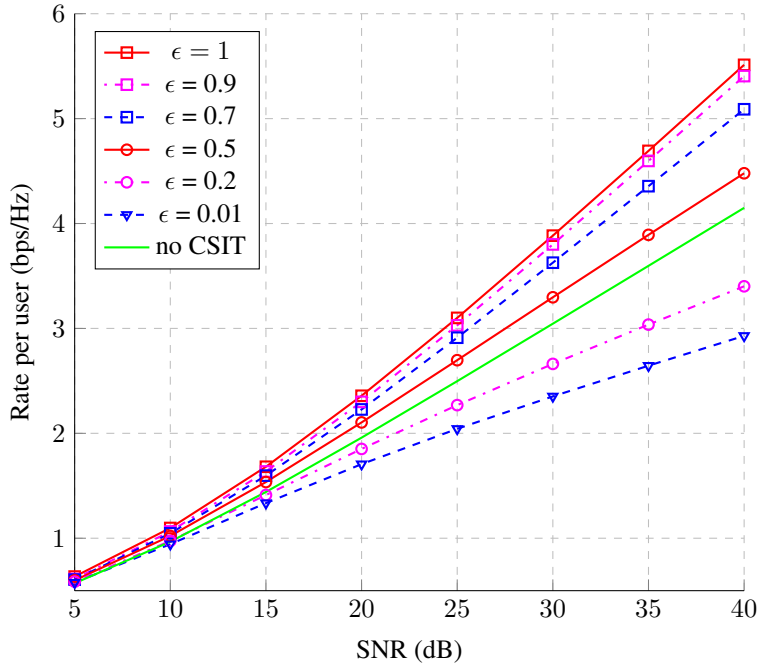  
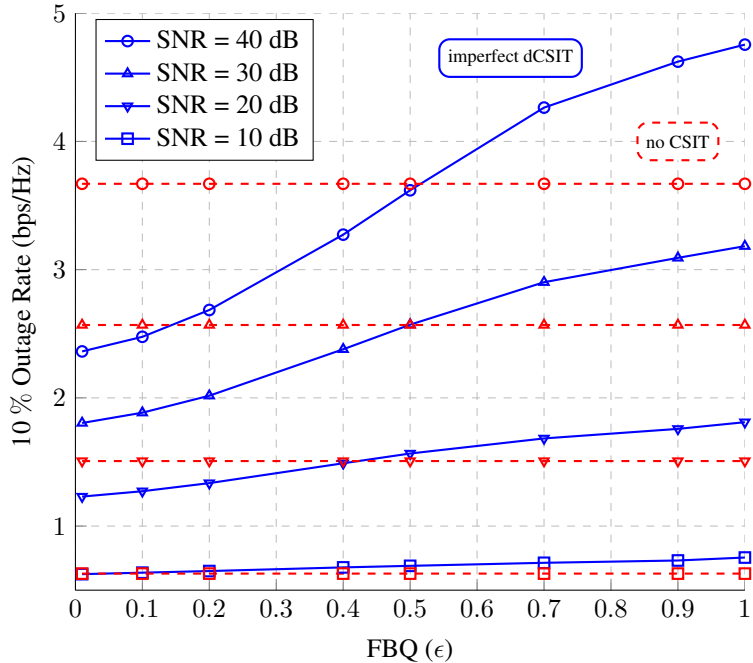
\begin{figure}[h] 
\begin{minipage}[]{\linewidth}
  \centering
%
%
%
%
\begin{tikzpicture}

\begin{axis}[%
view={-0}{90},
width=0.5\linewidth,
scale only axis,
xmin=0, xmax=1,
xlabel={FBQ ($\epsilon$)},
xmajorgrids,
ymin=0.5, ymax=5,
ylabel={10 $\%$ Outage Rate (bps/Hz)},
ylabel style={at={(0.06,0.5)}},
ymajorgrids,
axis lines*=left,
grid style={dashed},
legend style={at={(0.03,0.99)},anchor=north west,align=left}]
\textboxy{0.9}{4}{no CSIT}
\textbox{0.65}{4.65}{imperfect dCSIT}

\addplot [
color=blue,
solid,
mark=o,
thick,
mark options={solid}
]
coordinates{
 (0.01,2.36219011315729)(0.1,2.47602290667511)(0.2,2.68526633003756)(0.4,3.27297593742384)(0.5,3.61890586585158)(0.7,4.26386925534646)(0.9,4.62233485832222)(1,4.75506963844128) 
};
\addlegendentry{SNR = 40 dB};

\addplot [
color=blue,
solid,
mark=triangle,
thick,
mark options={solid}
]
coordinates{
 (0.01,1.80398226553993)(0.1,1.88431311476925)(0.2,2.01669921846795)(0.4,2.37897079749368)(0.5,2.57082743999752)(0.7,2.90198355358766)(0.9,3.09187217171847)(1,3.18314744032553) 
};
\addlegendentry{SNR = 30 dB};

\addplot [
color=blue,
solid,
mark=triangle,
thick,
mark options={solid,,rotate=180}
]
coordinates{
 (0.01,1.22996665398253)(0.1,1.27164386512507)(0.2,1.33491712362932)(0.4,1.48965623745885)(0.5,1.5659000832054)(0.7,1.68352111841608)(0.9,1.75793447223554)(1,1.81016120444152) 
};
\addlegendentry{SNR = 20 dB};

\addplot [
color=blue,
solid,
thick,
mark=square,
mark options={solid}
]
coordinates{
 (0.01,0.624976558990498)(0.1,0.636539942224714)(0.2,0.649143971999893)(0.4,0.677634095405013)(0.5,0.689776867112197)(0.7,0.713461210303841)(0.9,0.731488479939434)(1,0.754602273960632) 
};
\addlegendentry{SNR = 10 dB};

\addplot [
color=red,
dashed,
mark=o,
thick,
mark options={solid}
]
coordinates{
 (0.01,3.66945227369457)(0.1,3.66945227369457)(0.2,3.66945227369457)(0.4,3.66945227369457)(0.5,3.66945227369457)(0.7,3.66945227369457)(0.9,3.66945227369457)(1,3.66945227369457) 
};

\addplot [
forget plot,
color=red,
dashed,
mark=triangle,
thick,
mark options={solid}
]
coordinates{
 (0.01,2.56819174647074)(0.1,2.56819174647074)(0.2,2.56819174647074)(0.4,2.56819174647074)(0.5,2.56819174647074)(0.7,2.56819174647074)(0.9,2.56819174647074)(1,2.56819174647074) 
};

\addplot [
color=red,
dashed,
mark=triangle,
thick,
mark options={solid,,rotate=180}
]
coordinates{
 (0.01,1.5070259576246)(0.1,1.5070259576246)(0.2,1.5070259576246)(0.4,1.5070259576246)(0.5,1.5070259576246)(0.7,1.5070259576246)(0.9,1.5070259576246)(1,1.5070259576246) 
};

\addplot [
color=red,
dashed,
mark=square,
thick,
mark options={solid}
]
coordinates{
 (0.01,0.629269918867727)(0.1,0.629269918867727)(0.2,0.629269918867727)(0.4,0.629269918867727)(0.5,0.629269918867727)(0.7,0.629269918867727)(0.9,0.629269918867727)(1,0.629269918867727) 
};

\end{axis}
\end{tikzpicture}%
\end{minipage}
\caption{Outage rate vs FBQ for different SNR values.}
\label{fig:sims2}
\end{figure}                    

\newpage

\bibliographystyle{IEEEbib}
\bibliography{KUMIC}

\end{document}